\documentclass[twocolumn,english,aps,prl,twocolumn,superscriptaddress,showpacs,floatfix]{revtex4-1}
\usepackage{mathptmx}

\usepackage[T1]{fontenc}
\usepackage{amsmath}
\usepackage{amssymb}
\usepackage{graphicx}

\makeatletter

\providecommand{\tabularnewline}{\\}

\makeatother

\usepackage{babel}
\begin{document}


\title{Magnetic Interactions of the Centrosymmetric Skyrmion Material Gd$_{2}$PdSi$_{3}$}
\author{Joseph A. M. Paddison}
\email{paddisonja@ornl.gov}

\affiliation{Materials Science and Technology Division, Oak Ridge National Laboratory,
Oak Ridge, TN 37831, USA}
\author{Binod K. Rai}
\affiliation{Materials Science and Technology Division, Oak Ridge National Laboratory,
Oak Ridge, TN 37831, USA}
\affiliation{Savannah River National Laboratory, Aiken, South Carolina, 29808,
USA}
\author{Andrew F. May}
\affiliation{Materials Science and Technology Division, Oak Ridge National Laboratory,
Oak Ridge, TN 37831, USA}
\author{Stuart A. Calder}
\affiliation{Neutron Scattering Division, Oak Ridge National Laboratory, Oak Ridge,
Tennessee 37831, USA}
\author{Matthew B. Stone}
\affiliation{Neutron Scattering Division, Oak Ridge National Laboratory, Oak Ridge,
Tennessee 37831, USA}
\author{Matthias D. Frontzek}
\affiliation{Neutron Scattering Division, Oak Ridge National Laboratory, Oak Ridge,
Tennessee 37831, USA}
\author{Andrew D. Christianson}
\email{christiansad@ornl.gov}

\affiliation{Materials Science and Technology Division, Oak Ridge National Laboratory,
Oak Ridge, TN 37831, USA}
\begin{abstract}
The experimental realization of magnetic skyrmions in centrosymmetric
materials has been driven by theoretical understanding of how a delicate
balance of anisotropy and frustration can stabilize topological spin structures in applied magnetic fields. Recently, the centrosymmetric material Gd$_{2}$PdSi$_{3}$ was shown to host a field-induced skyrmion phase, but the skyrmion stabilization
mechanism remains unclear. Here, we employ neutron-scattering measurements
on an isotopically-enriched polycrystalline Gd$_{2}$PdSi$_{3}$ sample
to quantify the interactions that drive skyrmion formation. Our analysis
reveals spatially-extended interactions in triangular planes that are consistent with an RKKY
mechanism, and large ferromagnetic inter-planar magnetic interactions that
are modulated by the Pd/Si superstructure. The skyrmion phase emerges from a zero-field helical magnetic
order with magnetic moments perpendicular to the magnetic propagation
vector, indicating that the magnetic dipolar interaction plays a significant
role. Our experimental results establish an interaction space that can promote
skyrmion formation, facilitating identification and design of centrosymmetric skyrmion materials. 
\end{abstract}
\maketitle
Magnetic skyrmions are topologically-nontrivial spin textures with
potentially transformative applications in quantum computing and information
storage \citep{Tokura_2021,Fert_2017,Bogdanov_2020}. Skyrmions usually
occur in noncentrosymmetric magnets, in which they can be stabilized
by antisymmetric exchange interactions \citep{Muhlbauer_2009,Yu_2010}.
However, it was recently shown that skyrmions can be stabilized in
centrosymmetric systems by frustrated (competing) interactions \citep{Okubo_2012,Leonov_2015},
presenting the exciting prospects of higher skyrmion densities and
manipulation of chiral degrees of freedom by external fields \citep{Yu_2012,Yao_2020}.
While a small number of candidate centrosymmetric skyrmion materials
have been identified \citep{Kurumaji_2019,Hirschberger_2019,Gao_2020,Khanh_2020},
experimentally determining the magnetic interactions in such materials
remains a key challenge. Addressing this challenge is a prerequisite
for designing and manipulating skyrmion-based devices.

\begin{figure}
\centering{}\includegraphics{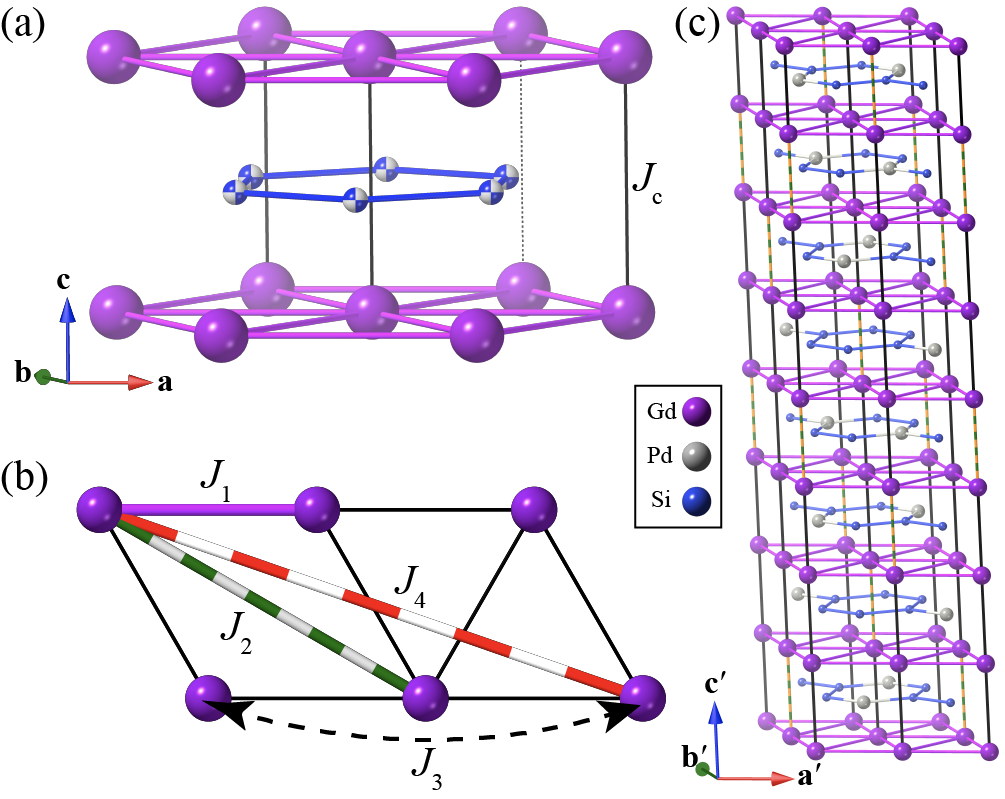}\caption{\label{fig:fig1} (a) High-symmetry crystal structure of Gd$_{2}$PdSi$_{3}$
(space group $P6/mmm$; $a=4.0618(1)$\,\AA, $c=4.0804(2)$\,\AA~at 25\,K, from our neutron diffraction data). (b) Magnetic interactions
within triangular Gd$^{3+}$ layers. (c) Proposed low-symmetry Pd/Si
superstructure showing ...ABCDBADC... stacking of PdSi$_{3}$ layers
($a^{\prime}=b^{\prime}=2a$, $c^{\prime}=8c$). The highest-symmetry
space group compatible with the superlattice ordering is $Fddd$ (see
SI). Black lines show inter-layer bonds with two Pd and four Si neighbors, and striped orange/green lines show inter-layer bonds with six Si neighbors.}
\end{figure}

The hexagonal material Gd$_{2}$PdSi$_{3}$ provides a rare
example of a skyrmion phase in a centrosymmetric system \citep{Kurumaji_2019}. In Gd$_{2}$PdSi$_{3}$, triangular layers
of magnetic Gd$^{3+}$ ions are separated by honeycomb PdSi$_{3}$ layers {[}Fig.~\ref{fig:fig1}(a){]}
\citep{Tang_2011}. A transition from the paramagnetic state occurs
at $T_{\mathrm{N}}=21$\,K to an incommensurate magnetic order with
propagation vector $\mathbf{q}=[q00]^{\ast}$ with $q\approx0.14$
\citep{Kurumaji_2019}. The observed $\mathbf{q}$ may be stabilized
by competition between ferromagnetic nearest-neighbor interactions
and antiferromagnetic further-neighbor interactions {[}Fig.~\ref{fig:fig1}(a,b){]}
\citep{Okubo_2012,Leonov_2015}. Application of a magnetic field below
$T_{\mathrm{N}}$ yields a giant topological Hall effect, signifying
a transition to a topologically-nontrivial skyrmion phase, which is
a triple-$\mathbf{q}$ structure formed by superposing magnetic helices
with $\mathbf{q}=[q00]^{\ast}$, $[0q0]^{\ast}$, and $[\bar{q}q0]^{\ast}$
\citep{Kurumaji_2019}. The bulk magnetic susceptibility follows a
Curie-Weiss law with spin $S=7/2$, $g=2$, and a ferromagnetic Weiss
temperature $\theta\approx30$\,K, indicating that Gd$^{3+}$ ions
possess spin-only local moments \citep{Kotsanidis_1990,Saha_1999,Zhang_2020}.
However, coupled electronic and spin correlations develop well above
$T_{\mathrm{N}}$, as indicated by a minimum in the resistivity at
$\sim$$2T_{\mathrm{N}}$ and a large negative magnetoresistance that
persists up to $\sim$$3T_{\mathrm{N}}$ \citep{Mallik_1998,Saha_1999,Zhang_2020}.

To explain spin textures in centrosymmetric systems such as Gd$_{2}$PdSi$_{3}$,
it is crucial to understand the system's underlying magnetic interactions.
The experimental observation of Fermi surface nesting with a wavevector
similar to $\mathbf{q}$ suggests the relevance of long-ranged RKKY
interactions \citep{Inosov_2009}, while a theoretical study indicates
that local exchange processes are also important \citep{Nomoto_2020}.
However, quantifying the interactions experimentally is a complex
problem, for three main reasons. First, the ordered magnetic structure
in zero applied field is not conclusively solved \citep{Kurumaji_2019,Zhang_2020,Moody_2021}.
Second, although the crystal structure may be approximately described
with a statistical distribution of Pd and Si, these atoms actually
form a superlattice that may affect exchange processes {[}Fig.~\ref{fig:fig1}(c){]}
\citep{Tang_2011}. Third, the large neutron absorption cross-section
of isotopically-natural Gd makes neutron-scattering experiments on large single
crystals challenging. So far, this has prevented the use of single-crystal
neutron-scattering experiments to understand the magnetic interactions
of Gd$_{2}$PdSi$_{3}$.

Here, we employ comprehensive neutron-scattering experiments on $^{160}$Gd$_{2}$PdSi$_{3}$
to obtain a model of its zero-field magnetic structure and interactions
that explains multiple experimental observations. We obtain the following
key results. First, magnetic interactions within triangular layers
are spatially extended and of competing sign, consistent with an RKKY
mechanism \citep{Wang_2020,Wang_2021} that is supported by a
comparison of our neutron-scattering results with published resistivity
data \citep{Saha_1999}. Second, ferromagnetic interactions between
layers are dominant, and strongly modulated by the Pd/Si superlattice.
Third, below $T_{\mathrm{N}}$, a helix with the spin plane perpendicular
to \textbf{$\mathbf{q}$} is the only structure consistent
with our neutron data and physical constraints, suggesting the magnetic dipolar interaction plays a significant role below $T_{\mathrm{N}}$
\citep{Utesov_2021}. Finally, we confirm that our interaction model
explains the skyrmion phase at small
applied magnetic fields below $T_{\mathrm{N}}$ \citep{Kurumaji_2019,Hirschberger_2020a}.
Our results provide a foundation for theoretical modeling and experimental
manipulation of spin textures in Gd$_{2}$PdSi$_{3}$.

\begin{figure}
\centering{}\includegraphics{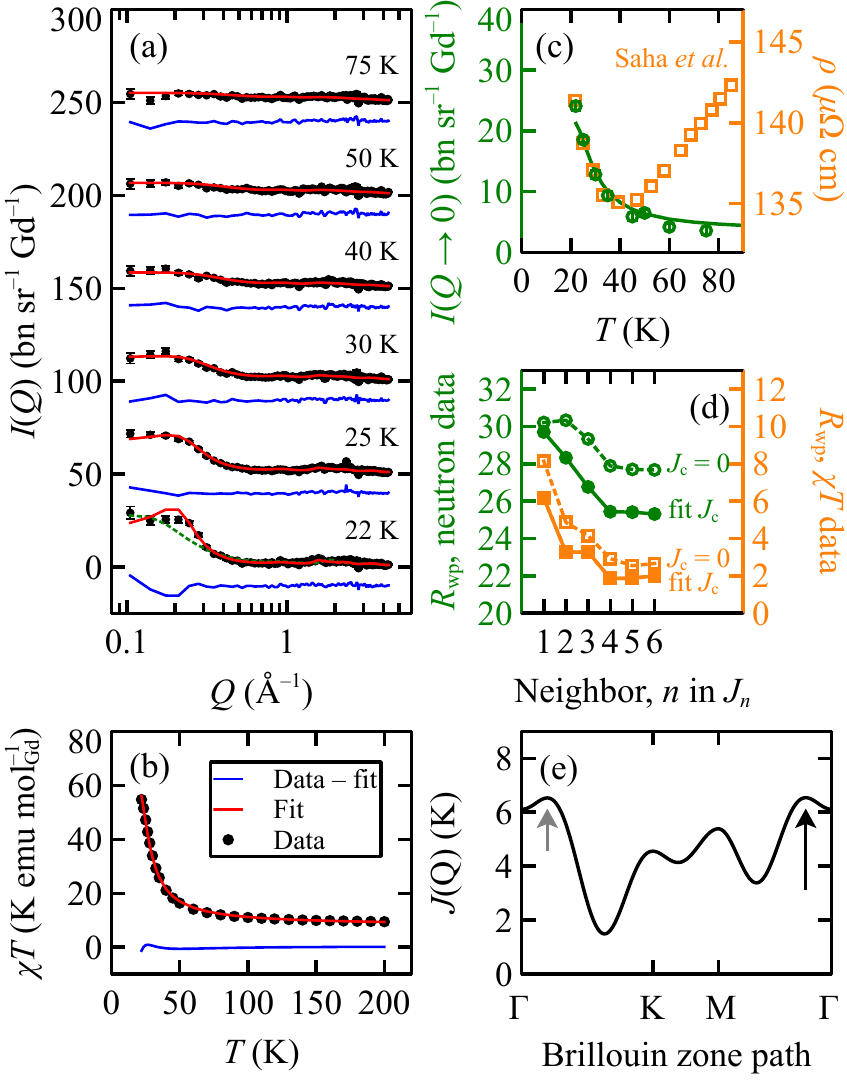}\caption{\label{fig:fig2}(a) Magnetic diffuse scattering above $T_{\mathrm{N}}$,
showing experimental data (black circles), model fits (red lines),
and data\,--\,fit (blue lines). Temperatures are labeled above each curve
and successive curves are shifted vertically by 50 bn\,sr$^{-1}$\,Gd$^{-1}$.
Data collected and fitted at $35$\,K, $45$\,K, and $60$\,K follow
the same trends and are omitted for clarity. The dotted green line
shows the $22$\,K fit with five intra-layer couplings
and $J_{\mathrm{c}}=0$. (b) Bulk magnetic susceptibility data and
fit (colors as above). (c) Comparison of $I(Q\rightarrow0)$ from neutron data 
(green circles, left axis) and magnetic susceptibility data
(solid green line, left axis) with published resistivity data from
Ref.~\citealp{Saha_1999} (orange squares, right axis). (d) Dependence
of goodness-of-fit metric $R_{\mathrm{wp}}$ for neutron data (green
circles, left axis) and susceptibility data (orange squares, right axis)
on the number of intra-layer neighbors, $n$ in $J_{n}$. Solid symbols show results when inter-layer coupling
$J_{\mathrm{c}}$ was fitted, and open symbols show results for $J_{\mathrm{c}}=0$.
(e) Dependence of $J(\mathbf{Q})$ along high-symmetry Brillouin zone
paths ($\Gamma=(000)$; $\mathrm{K}=(\frac{1}{3}\frac{1}{3}0)$; $\mathrm{M}=(\frac{1}{2}00)$).
Positions of global and local maxima in $J(\mathbf{Q})$ are
shown by long black and short gray arrows, respectively.}
\end{figure}

We prepared a polycrystalline sample of $^{160}$Gd$_{2}$PdSi$_{3}$
suitable for neutron measurements (mass $\sim$$0.8$ g) by arc melting.
Arc-melted samples were wrapped in Ta foil, sealed in a quartz tube
under a vacuum, and annealed at $800$\,C for one week. The sample
quality was confirmed by bulk magnetometry, which agrees with published
results, and by powder X-ray diffraction, which reveals broad superlattice
peaks consistent with $126(6)$\,\AA~domains of the superstructure
shown in Fig.~\ref{fig:fig1}(c) (see SI). To minimize neutron absorption,
the sample was 98.1\% enriched with $^{160}$Gd, and an annular sample
geometry was used for neutron diffraction and spectroscopy experiments,
which were performed using the HB-2A and SEQUOIA instruments at ORNL,
respectively. 

Figure~\ref{fig:fig2}(a) shows magnetic diffuse-scattering data
$I(Q)$ collected above $T_{\mathrm{N}}$ using HB-2A ($\lambda=2.4067$\,\AA).
The data are background-subtracted and placed in absolute intensity
units by normalization to the nuclear Bragg scattering. As the sample
is cooled below 40\,K, $I(Q)$ increases at small wavevectors, $Q\lesssim0.3$\,\AA$^{-1}$,
indicating the development of predominantly ferromagnetic short-range
correlations. Figure~\ref{fig:fig2}(b) shows that the bulk magnetic
susceptibility $\chi T$ exhibits a large upturn over the same temperature
range, as expected because $\chi T\propto I(\mathbf{Q}=\mathbf{0})$
at high temperature \citep{Lovesey_1987}. For RKKY interactions with
Fermi wave-vector $k_{\mathrm{F}}$, theory predicts an increase in
$I(Q\lesssim2k_{\mathrm{F}})$ as $T_{\mathrm{N}}$ is approached
from above, while a simultaneous enhancement of electron scattering
generates an upturn in the resistivity \citep{Wang_2016}. To test
this prediction, Fig.~\ref{fig:fig2}(c) compares the value of $I(Q\rightarrow0)$---obtained
from $\chi T$ and by averaging $I(Q)$ over $0.1\leq Q\leq0.3$\,\AA$^{-1}$---with
published resistivity measurements \citep{Saha_1999}. Both $I(Q\rightarrow0)$
and the resistivity shown an upturn at the same temperature ($\sim$$40$\,K),
in qualitative agreement with the RKKY prediction \citep{Wang_2016}.
This result suggests that RKKY interactions likely play a significant
role in Gd$_{2}$PdSi$_{3}$.

We quantify the magnetic interactions by analyzing
$I(Q)$ and $\chi T$ data measured at $T>T_{\mathrm{N}}$ within a Heisenberg model,
\begin{align*}
H_{\mathrm{ex}} & =-\frac{1}{2}\sum_{i,j}J_{ij}\mathbf{S}_{i}\cdot\mathbf{S}_{j},
\end{align*}
where $\mathbf{S}_{i}$ denotes a classical spin vector with position
$\mathbf{R}_{i}$ and length $\sqrt{S(S+1)}$, and the interaction
parameters $J_{ij}\in\{J_{1},J_{2},J_{3},J_{4},J_{\mathrm{c}}\}$ are shown
in Fig.~\ref{fig:fig1}(a,b). We make two simplifying assumptions
in this high-temperature analysis. First, we neglect non-Heisenberg
terms such as the magnetic dipolar interaction and single-ion anisotropy,
which have negligible effect above $T_{\mathrm{N}}$ because of their
small energy scales (see SI). Second, we assume the high-symmetry hexagonal structure,
neglecting a possible variation in $J_{ij}$ due to the Pd/Si superstructure.
Within reaction-field theory, the wavevector-dependent susceptibility
is then given by \citep{Logan_1995}
\[
\chi(\mathbf{Q})=\frac{\chi_{0}}{1-\chi_{0}[J(\mathbf{Q})-\lambda]},
\]
where $J(\mathbf{Q})=\sum_{j}J_{ij}\exp(\mathrm{i}\mathbf{Q}\cdot\mathbf{R}_{j})$, $\chi_{0}=S(S+1)/3$, $\lambda$ is obtained self-consistently
by enforcing that $\int_{\mathrm{BZ}}\chi(\mathbf{Q})\mathrm{d}\mathbf{Q}=S(S+1)$
\citep{Logan_1995}, and $I(Q)$ is calculated by spherically averaging $I(\mathbf{Q})=CT[f(Q)]^{2}\chi(\mathbf{Q})$,
where $C=0.1937$\,bn \citep{Lovesey_1987} and $f(Q)$ is the Gd$^{3+}$ magnetic form factor \citep{Brown_2004}.

\begin{table}
\centering{}%
\begin{tabular}{ccccc}
\hline 
$J_{\mathrm{c}}$ (K) & $J_{1}$ (K) & $J_{2}$ (K) & $J_{3}$ (K) & $J_{4}$ (K)\tabularnewline
\hline 
\hline 
$1.97(46)$ & $0.31(9)$ & $0.19(15)$ & $0.27(18)$ & $-0.21(5)$\tabularnewline
\hline 
\end{tabular}\caption{\label{tab:table1}Fitted values of magnetic interaction parameters.
Parameter uncertainties indicate $3\sigma$ confidence intervals.}
\end{table}

To identify the effective dimensionality of the interactions, we first
tested a two-dimensional model by setting $J_{\mathrm{c}}=0$ and
refining the intra-layer couplings $\left\{ J_{1},...,J_{n}\right\} $
to our $I(Q)$ and $\chi T$ data shown in Fig.~\ref{fig:fig2}(a)
and (b). This model does not represent the data well, even when $J_{n}$
up to $n=6$ are included {[}green line in Fig.~\ref{fig:fig2}(a)
for $22$\,K{]}. By contrast, also refining the inter-layer coupling
$J_{\mathrm{c}}$ substantially improves the refinement quality metric
$R_{\mathrm{wp}}$ {[}Fig.~\ref{fig:fig2}(d){]}, demonstrating that
the interactions are three-dimensional. To estimate the spatial extent
of the interactions, Fig.~\ref{fig:fig2}(d) shows the dependence
of $R_{\mathrm{wp}}$ on the number of $J_{n}$ fitted in addition
to $J_{\mathrm{c}}$. No significant improvement is obtained for $n>4$;
hence, our minimal model contains $\left\{ J_{1},J_{2},J_{3},J_{4},J_{\mathrm{c}}\right\} $.
The optimal parameter values from a global fit to $I(Q)$ and $\chi T$
data are given in Table~\ref{tab:table1}. Ferromagnetic $J_{\mathrm{c}}$
is dominant, while intra-layer interactions compete between antiferromagnetic
$J_{4}$ and shorter-range ferromagnetic couplings, reminiscent of
the RKKY interaction. Figure~\ref{fig:fig2}(e) shows the corresponding $J(\mathbf{Q})$, which is maximal at the propagation vector of the model, $\mathbf{q}_{\mathrm{calc}}\approx[0.12,0,0]^{\ast}$.
While $\mathbf{q}_{\mathrm{calc}}$ is smaller than the measured low-temperature
$\mathbf{q}\approx[0.14,0,0]^{\ast}$, the difference is plausible
because $\mathbf{q}$ decreases with increasing temperature below
$T_{\mathrm{N}}$ \citep{Kurumaji_2019}. Interestingly, a local $J(\mathbf{Q})$
maximum occurs along the $[110]^{*}$ direction with $<0.2$\%
energy difference from $J(\mathbf{q}_{\mathrm{calc}})$. Fermi-surface
measurements of Gd$_{2}$PdSi$_{3}$ show a nesting wavevector $\sim[\frac{1}{6}\frac{1}{6}0]^{*}$
\citep{Inosov_2009}, while Tb$_{2}$PdSi$_{3}$ exhibits short-range
magnetic ordering with this periodicity \citep{Frontzek_2007}, suggesting
the quasi-degeneracy of our model may be a generic feature of these
materials. Finally, we considered an alternative five-parameter model
containing two inter-layer and three intra-layer couplings.
While this model yields a comparable refinement of $I(Q)$ and $\chi T$
measurements, it does not agree well with inelastic neutron-scattering
data (see SI).

\begin{figure}
\centering{}\includegraphics{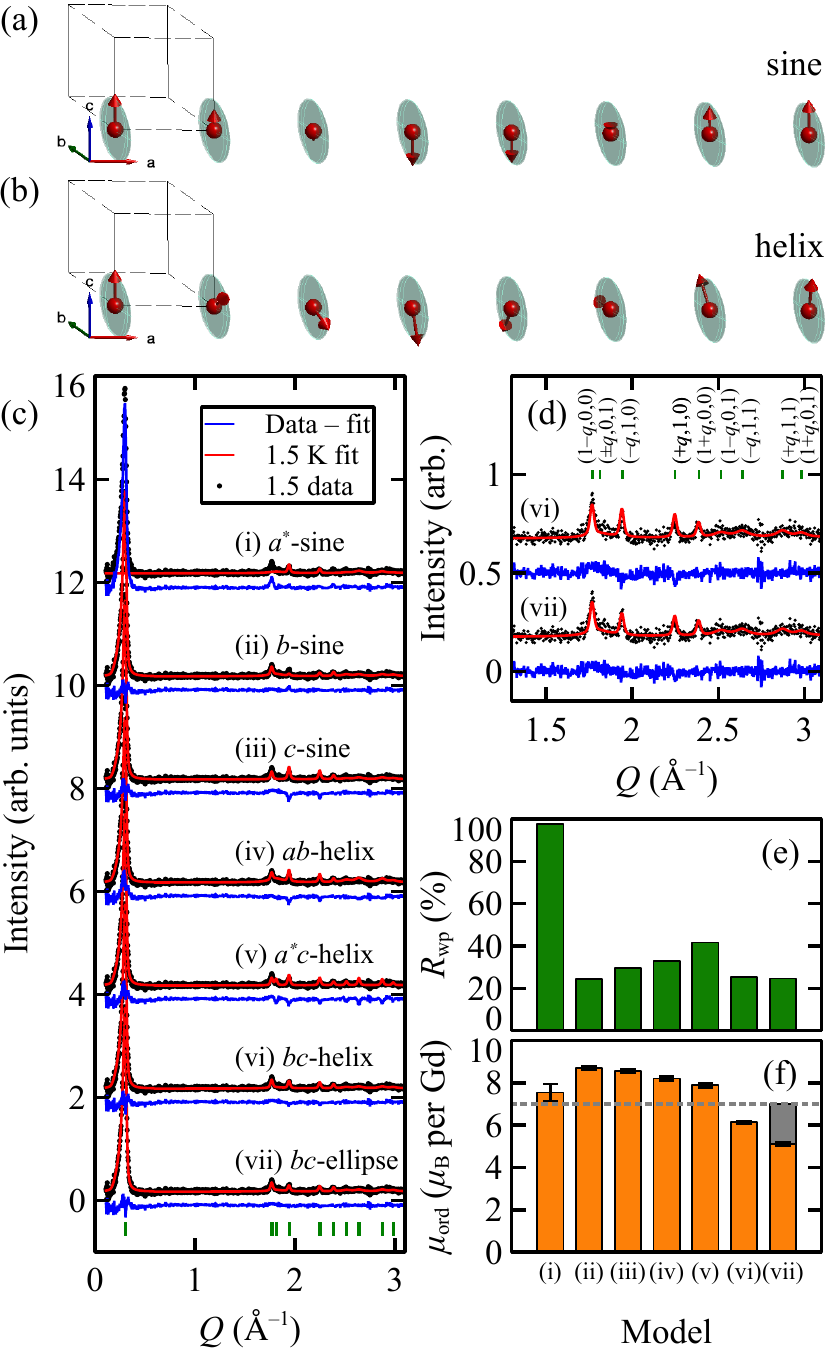}\caption{\label{fig:fig3}(a) Sinusoidal spin-density wave with
the spin axis perpendicular to $\mathbf{q}$. (b) ``Proper
screw'' helix with the spin plane perpendicular to $\mathbf{q}$.
(c) Magnetic diffraction data at 1.5\,K (black circles), model fits
(red lines), and data\,--\,fit (blue lines). Models are labeled
on each graph. (d) Magnetic diffraction data, fits and data\,--\,fit
(colors as above) on an expanded $Q$-axis scale for models (vi) and
(vii), showing broadening of peaks with $l\protect\neq0$ and
improved fit for the elliptical helix (vii) compared to the circular
helix (vi). (e) Goodness-of-fit metric $R_{\mathrm{wp}}$ for each
model (green bars). (f) Maximum refined value of the ordered magnetic
moment $\mu_{\mathrm{ord}}$ per Gd$^{3+}$ for each model (orange
bars). Parameter uncertainties represent $1\sigma$ confidence intervals.
For model (vii), $\mu_{\mathrm{ord}} \parallel \mathbf{b}$
is shown as a grey bar.}
\end{figure}

We now investigate the zero-field magnetic structure for $T<T_{\mathrm{N}}$.
Taking the hexagonal structure as the parent phase, there are three
magnetic irreducible representations (irreps) that correspond, respectively,
to sinusoidal modulations of the ordered magnetic moment $\mu_{\mathrm{ord}}$
along the orthogonal directions $\mathbf{a}^{\ast}$, $\mathbf{b}$,
and $\mathbf{c}$ {[}Fig.~\ref{fig:fig3}(a){]} \citep{Wills_2001}.
Alternatively, combining pairs of irreps yields helices with
$\mu_{\mathrm{ord}}$ in the $ab$, $a^{\ast}c$, or $bc$ plane,
as illustrated in Fig.~\ref{fig:fig3}(b). Both helical and sinusoidal
models have been proposed for the zero-field structure of Gd$_{2}$PdSi$_{3}$
\citep{Moody_2021,Nomoto_2020}. In addition, a triple-$\mathbf{q}$
meron-antimeron structure was proposed in Ref.~\citealp{Kurumaji_2019}.
In Fig.~\ref{fig:fig3}(c), we compare the Rietveld refinement for
each model with the measured magnetic diffraction pattern, obtained
as the difference between the 1.5\,K and 25\,K data. For each model,
Fig.~\ref{fig:fig3}(e) shows $R_{\mathrm{wp}}$, and Fig.~\ref{fig:fig3}(f)
shows the refined maximum value of $\mu_{\mathrm{ord}}$.
The $\mathbf{a}^{\ast}$-sine model (i), with spins $\mathbf{S} \parallel \mathbf{q}$,
would give zero intensity for the strong $(q00)$ magnetic peak, and
so is immediately ruled out. Of the remaining models, $\mathbf{b}$-sine
(ii), $bc$-helix (vi), and $bc$-ellipse (vii) structures yield similarly
high-quality refinements. The meron-antimeron structure has an identical
diffraction pattern to its single-$\mathbf{q}$ analog, the $bc$-helix,
and is not shown separately. The refined value of $\mu_{\mathrm{ord}}$
is a key discriminating factor, as any physical model must satisfy
the constraint that $\mathrm{max}(\mu_{\mathrm{ord}})\leq2S\mu_{\mathrm{B}}$
($=7.0\mu_{\mathrm{B}}$ for Gd$^{3+}$). This constraint rules out
the $\mathbf{b}$-sine model {[}Fig.~\ref{fig:fig3}(f){]}. It also
disfavors the meron-antimeron structure, for which $\mathrm{max}(\mu_{\mathrm{ord}})=\frac{3}{2}\mu_{\mathrm{ord}}^{\mathrm{helix}}$,
where $\mu_{\mathrm{ord}}^{\mathrm{helix}}=6.14(7)\thinspace\mu_{\mathrm{B}}$
is the refined ordered moment of the $bc$-helix. Thus, the key result
of our Rietveld analysis is that the only models yielding good fits
and reasonable $\mu_{\mathrm{ord}}$ values are ``proper screw''
helices with $\mathbf{S} \perp \mathbf{q}$, models
(vi) and (vii). The best refinement is for an elliptical helix with
$\mu_{\parallel \mathbf{c}}=5.13(7)\thinspace\mu_{\mathrm{B}}$, and $\mu_{\parallel \mathbf{b}}$
fixed to its maximum value of $7.0\thinspace\mu_{\mathrm{B}}$. Notably, the ordered moment is not fully
polarized as $\mu_{\mathrm{ord}}^{\mathrm{helix}}<2S\mu_{\mathrm{B}}$
at $1.5$\,K. Magnetic peaks are also selectively broadened compared
to nuclear peaks {[}Fig.~\ref{fig:fig3}(d){]}. Refinement of a quadratic-in-$l$
size-broadening term yields magnetic domain dimensions of $332(8)$\,\AA~in the $ab$-plane \emph{vs.} $27(2)$\,\AA~along $\mathbf{c}$,
which may be a consequence of disordered stacking of PdSi$_{3}$ layers.

\begin{figure}
\centering{}\includegraphics{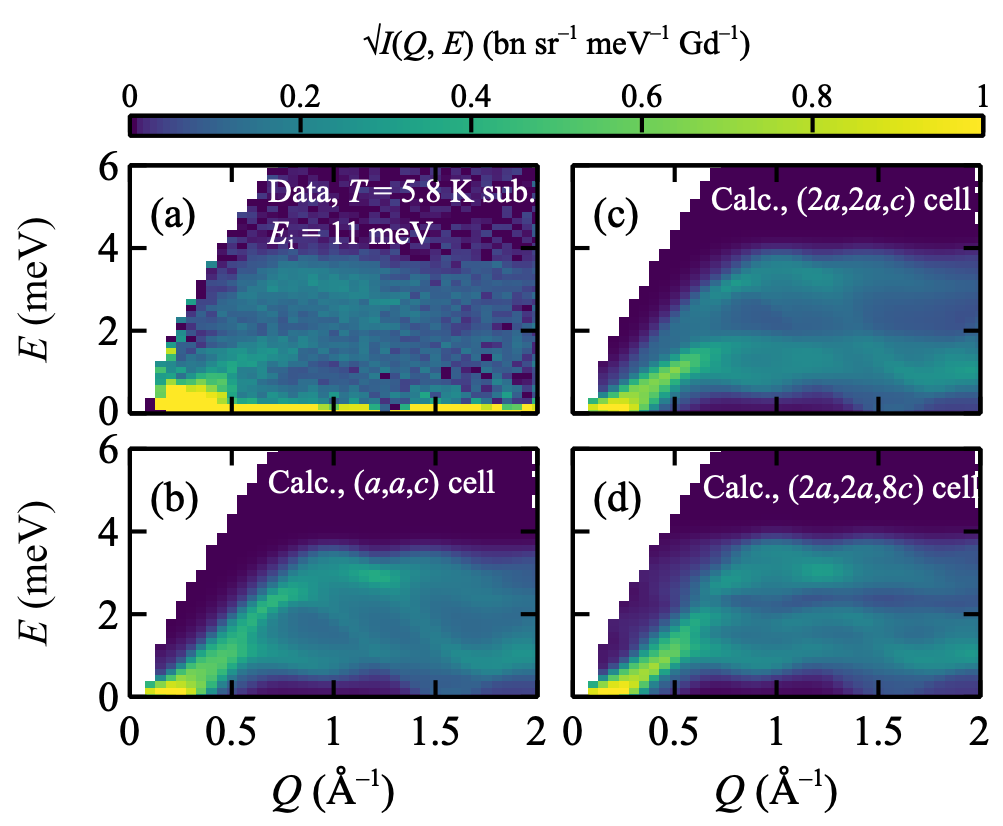}\caption{\label{fig:fig4}(a) Inelastic neutron-scattering data measured at
$T=5.8$\,K with $E_{i}=11$\,meV. Data are corrected for detailed
balance and diffuse scattering is subtracted. (b) Linear
spin-wave theory (LSWT) calculation for the interaction parameters
given in Table~\ref{tab:table1}. (c) LSWT calculation with $J_{\mathrm{c}}$
split by $\Delta=0.8$ (defined in the text) for Pd/Si superlattice
ordering with periodicity $(2a,2a,c)$. (d) LSWT calculation with
$J_{\mathrm{c}}$ split by $\Delta=0.8$ for Pd/Si superlattice ordering
with periodicity $(2a,2a,8c)$. }
\end{figure}

The magnetic excitation spectrum at $T\ll T_{\mathrm{N}}$
provides an exquisitely sensitive test of our model. Our inelastic
neutron-scattering data ($E_{i}=11$\,meV) show spin-wave excitations
at $T=5$\,K, superimposed on a diffuse magnetic background that
likely occurs because $\mu_{\mathrm{ord}}^{\mathrm{helix}}<2S\mu_{\mathrm{B}}$.
In Fig.~\ref{fig:fig4}(a), we show $I_{5\thinspace\mathrm{K}}^{\prime}=I_{5\thinspace\mathrm{K}}-[1-(\mu_{\mathrm{ord}}^{\mathrm{helix}}/2S\mu_{\mathrm{B}})^{2}]I_{25\thinspace\mathrm{K}}$,
which isolates the spin-wave contribution. Our data
show an overall bandwidth of approximately $4$\,meV.
For $E<4$\,meV, the spectrum has a broad energy dependence
with intensity minima for $0\lesssim E\lesssim1$\,meV and $2\lesssim E\lesssim3$\,meV.
Figure~\ref{fig:fig4}(b) shows the calculated spectrum
for the interaction parameters given in Table~\ref{tab:table1} and
a single-\textbf{$\mathbf{q}$} helical ground state, calculated within
linear spin-wave theory using the SpinW program \citep{Toth_2015}.
This model reproduces the overall bandwidth, but fails to explain
the intensity minimum for $2\lesssim E\lesssim3$\,meV. Attempts
to refine $\left\{ J_{1},J_{2},J_{3},J_{4},J_{\mathrm{c}}\right\} $
to the inelastic data also failed to reproduce this feature. To explain
our data, it was necessary to consider the effect of the Pd/Si superstructure
on $J_{\mathrm{c}}$. All proposed models of the Pd/Si superstructure
involve doubling the unit cell along $\mathbf{a}$ and $\mathbf{b}$,
such that 75\% of $J_{\mathrm{c}}$ bonds (notated $J_{\mathrm{c}+}$)
have four Si and two Pd neighbors, while the remaining $J_{\mathrm{c}}$
bonds (notated $J_{\mathrm{c}-}$) have six Si neighbors {[}Fig.~\ref{fig:fig1}(c){]}.
We assume the superstructure splits $J_{\mathrm{c}}$ by an amount
$\Delta J_{\mathrm{c}}$, such that $J_{\mathrm{c}+}=J_{\mathrm{c}}(1+\Delta/4)$ and $J_{\mathrm{c}-}=J_{\mathrm{c}}(1-3\Delta/4)$,
and neglect any splitting of the weaker interactions. For the $(2a,2a,8c)$
superstructure shown in Fig.~\ref{fig:fig1}(c), the stacking of
$J_{\mathrm{c}\pm}$ bonds is ...ABCDBADC... \citep{Tang_2011}, whereas
the $(2a,2a,c)$ superstructure considered in Ref.~\citealp{Nomoto_2020}
has ...AAA... stacking. Taking $\Delta=0.8$
with the $(2a,2a,c)$ superstructure reproduces the intensity minimum
for $2\lesssim E\lesssim3$ meV and yields good overall agreement
with our inelastic neutron-scattering data {[}Fig.~\ref{fig:fig4}(c){]}, without degrading the agreement with $I(Q)$ data above $T_\mathrm{N}$ (see SI). Taking
$\Delta=0.8$ with the $(2a,2a,8c)$ superstructure also generates
intensity minima, but yields worse agreement with our data {[}Fig.~\ref{fig:fig4}(d){]}.
Our results show that the Pd/Si superstructure strongly enhances $J_{\mathrm{c}}$
for bonds with Pd neighbors.

\begin{figure}
\centering{}\includegraphics{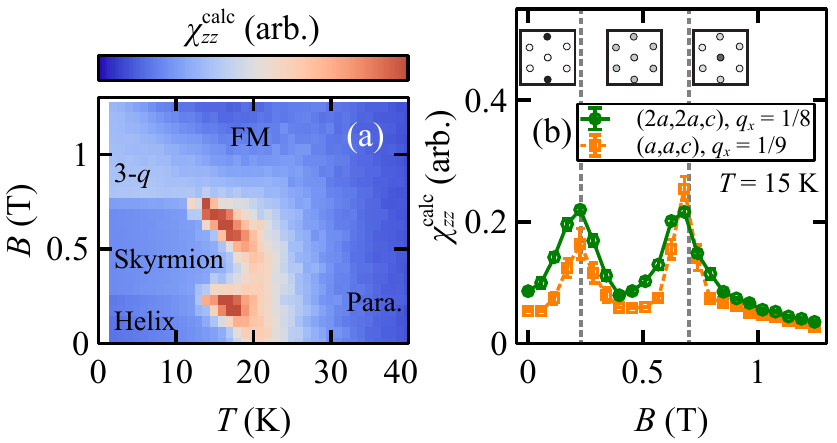}\caption{\label{fig:fig5}(a) Calculated magnetic susceptibility $\chi_{zz}^{\mathrm{calc}}$ for our interaction
model, obtained using Monte Carlo simulations. Results are shown for $q_{\mathrm{MC}}=\frac{1}{9}$ and a $9\times9\times9$
supercell of the hexagonal unit cell. (b) Calculated $\chi_{zz}^{\mathrm{calc}}$
at $T=15$\,K for the distorted $(2a,2a,c)$ supercell with $\Delta=0.8$
and $q_{\mathrm{MC}}=\frac{1}{8}$ (green circles), and the undistorted structure
with $\Delta=0$ and $q_{\mathrm{MC}}=\frac{1}{9}$ (orange squares). The calculated
magnetic diffraction patterns are for each phase are shown above,
for $B=0.11$, $0.40$, and $0.79$\,T (left to right). The values
of $B$ are scaled by the quantum correction factor $\sqrt{(S+1)/S}\approx1.134$.}
 
\end{figure}

We use extensive Monte Carlo simulations to calculate the phase diagram
of our model as a function of temperature $T$ and applied magnetic
field $\mathbf{B}\parallel\mathbf{c}$. The spin Hamiltonian is given
by
\[
H=H_{\mathrm{ex}}+g\mu_{\mathrm{B}}B\sum_{i}S_{i}^{z}+D\sum_{i>j}\frac{\mathbf{S}_{i}\cdot\mathbf{S}_{j}-3\left(\mathbf{S}_{i}\cdot\hat{\mathbf{r}}_{ij}\right)\left(\mathbf{S}_{j}\cdot\hat{\mathbf{r}}_{ij}\right)}{\left(r_{ij}/r_{1}\right)^{3}},
\]
where, to stabilize helical ordering with $\mathbf{S}\perp\mathbf{q}$,
we include the magnetic dipolar interaction that has magnitude
$D=0.037$\,K at the nearest-neighbor distance $r_{1}$ \citep{Utesov_2021,Utesov_2021a}.
To minimize finite-size effects, we constrain the interactions to
stabilize $\mathbf{q}_{\mathrm{MC}}=[q_{\mathrm{MC}}00]^{\ast}\approx\mathbf{q}_{\mathrm{calc}}$,
with commensurate $q_{\mathrm{MC}}=\frac{1}{8}$ or $\frac{1}{9}$. The calculated
magnetic susceptibility $\chi_{zz}^{\mathrm{calc}}(B,T)$ is shown in Fig.~\ref{fig:fig5}(a),
and reveals both similarities and differences with experiment \citep{Spachmann_2021,Hirschberger_2020a}. In agreement
with experiment, we find $T_{\mathrm{N}}^{\mathrm{calc}}\approx20$\,K,
and below $T_{\mathrm{N}}$, a transition from a helical to a skyrmion
phase at $B\approx0.25$\,T. At larger $B$, a further transition
occurs to a topologically-trivial triple-\textbf{q} phase previously
identified using mean-field theory \citep{Utesov_2021}. The single-$\mathbf{q}$
\emph{vs.} triple-$\mathbf{q}$ nature of each phase is revealed by
its calculated magnetic diffraction pattern {[}insets in Fig.~\ref{fig:fig5}(b){]}.
The behavior is not qualitatively affected by the splitting of $J_{\mathrm{c}}$, or by the precise value of $q_{\mathrm{MC}}$ {[}Fig.~\ref{fig:fig5}(b){]}.
While our model shows good agreement with experiment at small applied
fields, it does not explain the large increase in saturation field
on cooling the sample ($B_{\mathrm{sat}}\approx8$\,T at $2$\,K
\citep{Hirschberger_2020a}) or the presence of magnetic transitions
for $B>1$\,T \citep{Spachmann_2021}. These differences motivate
further theoretical work to understand the role of non-Heisenberg interactions.

Our neutron-scattering results provide an experimental understanding
of the magnetic interactions in Gd$_{2}$PdSi$_{3}$ and clarify its
zero-field magnetic structure. Our identification of the hierarchy of
energy scales will facilitate the development of experiments
to manipulate spin textures in Gd$_{2}$PdSi$_{3}$. Notably, our
interaction model explains key aspects of the experimental behavior
without invoking biquadratic or multi-spin interactions \citep{Hayami_2021b}.
However, the spin dynamics can only be understood by accounting for
the Pd/Si superstructure, suggesting it is important to include this in models. We anticipate that this model of the skyrmion
stabilization mechanism in Gd$_{2}$PdSi$_{3}$ will facilitate design
and identification of new centrosymmetric skyrmion hosts, including
in materials where large single-crystal samples are unavailable or
unsuitable for neutron-scattering measurements.

\begin{acknowledgments}
We are grateful to Cristian Batista, Matthew Cliffe, Randy Fishman, Shang Gao, and Stephen
Nagler for valuable discussions.
This work was supported by the U.S. Department of Energy, Office of
Science, Basic Energy Sciences, Materials Sciences and Engineering
Division. This research used resources at the High Flux Isotope Reactor
and Spallation Neutron Source, DOE Office of Science User Facilities
operated by the Oak Ridge National Laboratory.
\end{acknowledgments}


\end{document}